\documentclass[12pt]{article}

\usepackage{graphicx}

\begin{document}

\begin{center}
{\bf Nonlinear Electromagnetic Fields As a Source of Universe Acceleration} \\
\vspace{5mm} S. I. Kruglov
\footnote{E-mail: serguei.krouglov@utoronto.ca}

\vspace{3mm}
\textit{Department of Chemical and Physical Sciences, University of Toronto,\\
3359 Mississauga Road North, Mississauga, Ontario L5L 1C6, Canada} \\
\vspace{5mm}
\end{center}

\begin{abstract}
A model of nonlinear electromagnetic fields with a dimensional parameter $\beta$ is proposed. From PVLAS experiment the bound on the parameter $\beta$ was obtained. Electromagnetic fields are coupled with the gravitation field and we show that the universe accelerates due to nonlinear electromagnetic fields. The magnetic universe is considered and the stochastic magnetic field is a background. After inflation the universe decelerates and approaches to the radiation era. The range of the scale factor, when the causality of the model and a classical stability take place, was obtained. The spectral index, the tensor-to-scalar ratio, and the running of the spectral index were estimated which are in approximate agreement with the PLANCK, WMAP, and BICEP2 data.
\end{abstract}

\section{Introduction}

Last years nonlinear electromagnetic fields are of interest in cosmological models \cite{Garcia}, \cite{Camara}, \cite{Elizalde}, \cite{Novello}, \cite{Novello1}, \cite{Vollick}, \cite{Kruglov3}.
The Standard Cosmological Model (SCM) does not solve the problems of the initial singularity and the current universe acceleration. Thus, SCM is based on Big Bang scenario that leads to curvature invariants singularities.
But the initial singularities in the early universe can be avoided if one uses nonlinear modified Maxwell's equations \cite{Novello1}. Models of the magnetic universe show the absence of the initial singularity because the electromagnetic field in modified nonlinear electrodynamics (NLED) is strong in the early universe.
In some models of NLED the correspondence principle, that for weak electromagnetic fields the Lagrangian density approaches to the Maxwell Lagrangian density, is broken \cite{Novello1}.
In this paper we investigate the cosmological model, based on new type of NLED, that contains a stochastic magnetic background with a non-vanishing $< B^2 >$. For our model the correspondence principle takes place. In addition, the model avoids initial singularities. There are some models that can solve the problems of initial singularities without introducing NLED. We mention the model of pre-big-bang universe in superstring cosmology introducing a scalar field \cite{Gasperini} and the ekpyrotic universe exploring branes \cite{Turok}. In NLED-cosmology, based on classical physics, there are not new degrees of freedom such as scalars or branes. Therefore, in our opinion, cosmology that uses NLED possesses some attractive features.

The proposal of the paper is to show that nonlinear electromagnetic radiation which is described by modified Maxwell's equations is a source of the universe inflation. To explain the universe acceleration one can introduce the cosmological constant, $\Lambda$, in Einstein's equation or to introduce a scalar field with some potential function or to modify the gravity theory with the help of some function $F(R)$. There is a problem to explain the smallness of the $\Lambda$ compared to the vacuum energy.
In the case of scalar-tensor theory there are many potentials that lead to inflation and universe acceleration. In the modified gravity models the choice of functions $F(R)$ is not unique \cite{Capozziello}, \cite{Nojiri}, \cite{Bamba}, \cite{Kruglov4}, \cite{Kruglov5}, \cite{Kruglov6}, \cite{Kruglov7}, \cite{Kruglov8}, \cite{Kruglov9}. NLED coupled to the gravitation field may produce negative pressure and can lead to inflation of the universe \cite{Garcia}, \cite{Camara}, \cite{Elizalde}, \cite{Novello1}, \cite{Novello}, \cite{Vollick}, \cite{Kruglov3}.
In this paper we explore a new model of NLED coupled to gravity and a stochastic magnetic field drives inflation of the universe. The source of a stochastic magnetic field could be plasma fluctuations \cite{Lemoine}, \cite{Lemoine1}.
Thus, there are the stochastic fluctuations of the electromagnetic field in a relativistic electron-positron plasma \cite{Lemoine}. The scenario is to generate a primordial magnetic field from thermal fluctuations in the pre-recombination plasma. Magnetic fluctuations are sustained by plasma before the epoch of Big Bang nucleosynthesis (BBN) and the early universe was filled by a strong low-frequency random magnetic field during the early stage of the radiation-dominated era. Indeed, magnetic fields of the order of $B=10^{-6}$ G exist on scales of a few Kpc in our galaxy and other spiral galaxies \cite{Kronberg}. Such magnetic fields have the primordial origin and can be explained by the galactic dynamo theory (a mechanism transferring angular momentum energy into magnetic energy). However, the galactic dynamo theory requires the existence of weak seed fields. For successful dynamo amplification a seed field of the order of $B_{seed}=10^{-19}$ G is needed at the epoch of the galaxy formation. It is possible to generate seed magnetic fields by thermal fluctuations in the primordial plasma. Long wavelength fluctuations can reconnect and redistribute the magnetic energy over larger scales. A new scenario for the creation of galactic magnetic fields was proposed \cite{Dolgov}.

The nonlinear electrodynamics proposed is an effective model of electromagnetic fields that holds for very strong fields and at the weak field limit leads to Maxwell electrodynamics. As a result, without the modification of General Relativity (GR) the model explains the universe inflation. In the early time of the universe evolution the electromagnetic and gravitational fields are very strong and quantum corrections have to be taken into consideration \cite{Jackson} leading to NLED.

The goal of the paper is to describe inflation with the help of proposed model of NLED coupled to gravity.
Previously we have considered other models of NLED in \cite{Kruglov}, \cite{Krug}, \cite{Krug1}, \cite{Krug2}.
which have some differences compared to the current model. The model described in \cite{Kruglov3}
also leads to the acceleration of the universe. In addition to the previous work we have
showed that the current model describes the spectral index, the tensor-
to-scalar ratio, and the running of the spectral index in approximate agreement with the
PLANCK, WMAP, and BICEP2 data. The causality and a classical stability in this model also
take place at different periods of the universe evolution as compared with \cite{Kruglov3}. The dependence of the scale factor on the time, that describes the universe evolution, is different in the current model.

The paper is organized as follows. In section 2 a new model of nonlinear electromagnetic fields with a dimensional parameter $\beta$ is proposed. We obtain the energy-momentum tensor having non-vanishing trace. The field equations are written in the form of the Maxwell equations where the electric permittivity and magnetic permeability depend on electromagnetic fields. We obtain the bound on the coefficient $\beta$ from the PVLAS experiment. The NLED coupled to gravity is studied in Sec. 3. We investigate the magnetic universe with the background of stochastic magnetic fields. We show that there are no singularities of the energy density, pressure, the Ricci scalar and the Ricci tensor squared. It is demonstrated the universe inflation for strong magnetic fields. In Sec. 4 the universe evolution is studied. We solve Friedmann's equation and the function of the scale factor on the time is obtained. The range of the scale factor, when the causality of the model and a classical stability hold, was obtained. We have evaluated the spectral index $n_s$, the tensor-to-scalar ratio $r$, and the running of the spectral index $\alpha_s$ that are in approximate agreement with the PLANK, WMAP, and BICEP2 data. Conclusion is made in Sec. 5.

The units with $c=\hbar=\varepsilon_0=\mu_0=1$ and the metric $\eta=\mbox{diag}(-,+,+,+)$ are used.

\section{Nonlinear electromagnetic fields}

We propose here the model of nonlinear electromagnetic fields with the Lagrangian density
\begin{equation}
{\cal L} = -{\cal F}\exp(-\beta{\cal F}),
\label{1}
\end{equation}
where $F_{\mu\nu}$ is the field strength tensor, ${\cal F}=(1/4)F_{\mu\nu}F^{\mu\nu}=(\textbf{B}^2-\textbf{E}^2)/2$, and $\beta{\cal F}$ is dimensionless. The correspondence principle takes place because at ${\cal F}\rightarrow 0$ we have ${\cal L}\rightarrow -{\cal F}$.
Another variant of exponential electrodynamics was considered in \cite{Hendi}.
One can obtain the symmetric energy-momentum tensor by varying the
action with respect to the metric tensor \cite{Birula} so that
\begin{equation}
T^{\mu\nu}=H^{\mu\lambda}F^\nu_{~\lambda}-g^{\mu\nu}{\cal L},
\label{2}
\end{equation}
with
\begin{equation}
H^{\mu\lambda}=\frac{\partial {\cal L}}{\partial F_{\mu\lambda}}=\frac{\partial {\cal L}}{{\partial\cal F}}F^{\mu\lambda}=\left(\beta {\cal F}-1\right)\exp(-\beta{\cal F})F^{\mu\lambda}.
\label{3}
\end{equation}
The symmetric energy-momentum tensor found from Eqs. (2),(3) is given by
\begin{equation}
T^{\mu\nu}=\exp(-\beta{\cal F})\left[\left(\beta {\cal F}-1\right)F^{\mu\lambda}F^\nu_{~\lambda}+g^{\mu\nu}{\cal F}\right].
\label{4}
\end{equation}
From Eq. (4) we obtain the trace of the energy-momentum tensor
\begin{equation}
{\cal T}\equiv T_{\mu}^{~\mu}=4\beta {\cal F}^2\exp(-\beta{\cal F}).
\label{5}
\end{equation}
At weak fields or at $\beta\rightarrow 0$, we arrive at classical electrodynamics, ${\cal L}\rightarrow -{\cal F}$ and the trace of the energy-momentum tensor (5) vanishes, ${\cal T}\rightarrow 0$. In general, $\beta\neq 0$, and the non-zero energy-momentum tensor trace leads to the violation of the scale invariance. In any variants of nonlinear electrodynamics with the dimensional parameter the scale invariance is broken and the divergence of the dilatation current is nonzero, $\partial_\mu D^\mu={\cal T}$ ($D_\mu=x_\nu T_{\mu}^{~\nu}$).
We obtain the electric displacement field from the relation $\textbf{D}=\partial{\cal L}/\partial \textbf{E}$,
\begin{equation}
\textbf{D}=\left(1-\beta {\cal F}\right)\exp(-\beta{\cal F})\textbf{E},
\label{6}
\end{equation}
and from the definition $\textbf{D}=\varepsilon \textbf{E}$ we find the electric permittivity $\varepsilon$:
\begin{equation}
\varepsilon=\left(1-\beta {\cal F}\right)\exp(-\beta{\cal F}).
\label{7}
\end{equation}
With the help of the equality $\textbf{H}=-\partial{\cal L}/\partial \textbf{B}$ one can obtain the magnetic field
\begin{equation}
\textbf{H}= \left(1-\beta {\cal F}\right)\exp(-\beta{\cal F})\textbf{B}.
\label{8}
\end{equation}
By virtue of the relation $\textbf{B}=\mu \textbf{H}$, the magnetic permeability is given by $\mu=1/\varepsilon$. We find the relation $\textbf{D}\cdot\textbf{H}=\varepsilon^2\textbf{E}\cdot\textbf{B}$ from Eqs. (6),(8) showing that $\textbf{D}\cdot\textbf{H}\neq\textbf{E}\cdot\textbf{B}$ and, therefore, the dual symmetry is broken \cite{Gibbons} in the model proposed.
From the Lagrangian density (1) we obtain the field equations $\partial_\mu\left[(1-\beta{\cal F})\exp(-\beta{\cal F})F^{\mu\nu}\right]=0$, which can be represented, with the aid of Eqs. (6),(8), in the form of the first pair of Maxwell's equations
\begin{equation}
\nabla\cdot \textbf{D}= 0,~~~~ \frac{\partial\textbf{D}}{\partial
t}-\nabla\times\textbf{H}=0.
\label{9}
\end{equation}
The second pair of the Maxwell equations follows from the equation $\partial_\mu \widetilde{F}^{\mu\nu}=0$, with $\widetilde{F}_{\mu\nu}$ being the dual tensor, that is the consequence of the Bianchi identity
\begin{equation}
\nabla\cdot \textbf{B}= 0,~~~~ \frac{\partial\textbf{B}}{\partial
t}+\nabla\times\textbf{E}=0.
\label{10}
\end{equation}
The electric permittivity $\varepsilon$ and the magnetic permeability $\mu=1/\varepsilon$ depend on the electromagnetic fields $\textbf{E}$, $\textbf{B}$, and as a result, Eqs. (6), (8), (9), (10) represent the nonlinear equations of the electromagnetic fields. Last years different models of nonlinear electrodynamics attract attention due to some interesting effects discovered in such models \cite{Hendi1}, \cite{Kruglov}, \cite{Krug}, \cite{Krug1}, \cite{Krug2}.

\subsection{Vacuum birefringence}

The foundations of classical electrodynamics in flat spacetime were tested in experiments measuring vacuum birefringence: the BMV (Bir\'{e}fringence Magn\'{e}tique du Vide) experiment \cite{Battesti}, \cite{Rizzo}, the PVLAS (Polarizzazione del Vuoto con LASer) experiment \cite{Zavattini}, \cite{Valle} and the QA (QED vacuum birefringence and Axion search) experiment \cite{Chen}, \cite{Mei}. The effect of vacuum birefringence takes place in QED due to quantum corrections described by the Heisenberg-Euler effective Lagrangian \cite{Heisenberg} (see also \cite{Adler1}, \cite{Biswas}). The phenomenon of vacuum birefringence is absent in classical electrodynamics and in Born-Infeld (BI) \cite{Born} electrodynamics but in generalized BI electrodynamics with two parameters \cite{Kruglov1} the effect of vacuum birefringence takes place. Now we describe the vacuum birefringence in NLED based on the Lagrangian density (1). At $\beta {\cal F}\ll 1$ the Taylor series leads, by virtue of Eq. (1), to the Lagrangian density
\begin{equation}
{\cal L} = -{\cal F}\left[1-\beta{\cal F}+{\cal O}\left((\beta{\cal F})^2\right)\right].
\label{11}
\end{equation}
The effect of vacuum birefringence, when the external constant magnetic induction field is present, was investigated in \cite{Kruglov2} for the model with the Lagrangian density
\begin{equation}
{\cal L} = \frac{1}{2}\left(\textbf{E}^2-\textbf{B}^2\right)+a\left(\textbf{E}^2-\textbf{B}^2\right)^2
+b\left(\textbf{E}\cdot\textbf{B}\right)^2.
\label{12}
\end{equation}
By comparing Eqs. (11),(12), we make a conclusion that up to ${\cal O}\left((\beta{\cal F})^2\right)$ coefficients  are $a=\beta/4$, $b=0$ for the model under consideration. It was shown in \cite{Kruglov2} that for the model (12) the indexes of refraction $n_\perp$, $n_\|$ for two polarizations, perpendicular and parallel to the external magnetic induction field $\bar{B}$, are given by
\begin{equation}
n_\perp=1+4a\bar{B}^2,~~~~n_\|=1+b\bar{B}^2.
\label{13}
\end{equation}
Thus, for our model we have $n_\perp=1+\beta\bar{B}^2$, $n_\|=1$.
As a result, the phase velocities $v_\perp=1/n_\perp<1$, $v_\|=1/n_\|=c=1$ depend on the polarization of the electromagnetic wave, as compared to the direction of the external magnetic field, and we have the effect of vacuum birefringence.
According to the Cotton-Mouton (CM) effect \cite{Battesti} the difference in the indexes of refraction is
\begin{equation}
\triangle n_{CM}=n_\|-n_\perp=k_{CM}\bar{B}^2.
\label{14}
\end{equation}
For our model (1) the Cotton-Mouton coefficient is equal approximately to $k_{CM}=-\beta$.
The bounds on coefficient $k_{CM}$ from BMV and PVLAS experiments are
\[
k_{CM}=(5.1\pm 6.2)\times 10^{-21} \mbox {T}^{-2}~~~~~~~~~~(\mbox {BMV}),
\]
\begin{equation}
k_{CM}=(4\pm 20)\times 10^{-23} \mbox {T}^{-2}~~~~~~~~~~~~(\mbox {PVLAS}).
\label{15}
\end{equation}
Thus, the lower bound on the parameter from PVLAS experiment is $\beta=(-4\pm 20)\times 10^{-23} \mbox {T}^{-2}$.
Within QED, taking into account quantum corrections, the Cotton-Mouton coefficient is given by $k_{CM}\approx 4.0\times 10^{-24} \mbox {T}^{-2}$ \cite{Rizzo}. It should be noted that nonlinear effects play important role when the invariant is $\beta {\cal F}=\beta B^2/2\approx 1$ and we have non-Maxwell's electrodynamics. If one takes the value $\beta \approx 10^{-24} \mbox {T}^{-2}$ then nonlinear terms will have effects when the magnetic induction fields are in the order of $10^{12}$ T. Such strong magnetic fields may take place in the early universe.

\section{Cosmology}

It should be mentioned that electromagnetic fields play an important role in cosmology \cite{Dolgov}.
Let us consider the electromagnetic fields described by the Lagrangian density (1) coupled with the gravitation fields. Then the action of the gravitational and nonlinear electromagnetic fields (1), implying that NLED is a source of gravity, is given by
\begin{equation}
S=\int d^4x\sqrt{-g}\left[\frac{1}{2\kappa^2}R+ {\cal L}\right],
\label{16}
\end{equation}
where $R$ is the Ricci scalar, $\kappa^{-1}=M_{Pl}$, $M_{Pl}$ is the reduced Planck mass. After varying action (16) one finds the Einstein and electromagnetic field equations \footnote{The sign minus in the right side of Eq. (17) is due to our definition of the energy momentum tensor (4) and the Ricci scalar.}
\begin{equation}
R_{\mu\nu}-\frac{1}{2}g_{\mu\nu}R=-\kappa^2T_{\mu\nu},
\label{17}
\end{equation}
\begin{equation}
\partial_\mu\left[\sqrt{-g}(\beta{\cal F}-1)\exp(-\beta{\cal F})F^{\mu\nu}\right]=0.
\label{18}
\end{equation}
In the flat spacetime Eq. (18) is equivalent to the Maxwell equations (6), (8), (9), (10).
Let us consider homogeneous, isotropic and spatially flat Friedmann-Robertson-Walker (FRW) metric with the line element
\begin{equation}
ds^2=-dt^2+a(t)^2\left(dx^2+dy^2+dz^2\right),
\label{19}
\end{equation}
where $a(t)$ is a scale factor.
Here the electromagnetic fields are the stochastic background and we assume that the wavelength of electromagnetic waves is smaller than the curvature. To produce the isotropy of the FRW spacetime
we use the average of the electromagnetic fields \cite{Tolman}. Then
electromagnetic fields averaged possess the properties
\[
<\textbf{E}>=<\textbf{B}>=0,~~~~<E_iB_j>=0,
\]
\begin{equation}
<E_iE_j>=\frac{1}{3}E^2g_{ij},~~~~<B_iB_j>=\frac{1}{3}B^2g_{ij}.
\label{20}
\end{equation}
We suppose averaging over a volume that is larger than the radiation
wavelength and smaller as compared to the spacetime curvature. For simplicity we omit the brackets $<>$ in that follows. One can obtain the energy density $\rho$ and the pressure $p$ from the relations
\begin{equation}
\rho=-{\cal L}-E^2\frac{\partial {\cal L}}{\partial{\cal F}}=\exp(-\beta{\cal F})\left[{\cal F}+E^2(1-\beta{\cal F})\right],
\label{21}
\end{equation}
\begin{equation}
p={\cal L}+\frac{E^2-2B^2}{3}\frac{\partial {\cal L}}{\partial{\cal F}}=
-\exp(-\beta{\cal F})\left[{\cal F}+\frac{E^2-2B^2}{3}(1-\beta{\cal F})\right].
\label{22}
\end{equation}
Eqs. (21),(22) can also be obtained from the energy-momentum tensor (4).
The Friedmann equation follows from the Einstein equation (17) and FRW metric (19),
\begin{equation}
3\frac{\ddot{a}}{a}=-\frac{\kappa^2}{2}\left(\rho+3p\right).
\label{23}
\end{equation}
Dots over the variables give the derivatives with respect to the cosmic time $t$.
If $\rho + 3p < 0$ an acceleration of the universe takes place. It was shown in \cite{Lemoine1} that the conducting fluid and the electric fields are screened because of the charged primordial plasma. Therefore, we investigate the magnetic universe when $E = 0$. From Eqs. (21),(22) we find
\begin{equation}
\rho+3p=B^2(1-\beta B^2)\exp(-\beta B^2/2).
\label{24}
\end{equation}
It follows from Eq. (24) that the maximum acceleration occurs at $\beta B^2=(5+\sqrt{17})/2$ and the maximum deceleration occurs at $\beta B^2=(5-\sqrt{17})/2$.
The plot of the function $\beta(\rho+3p)$ versus $\beta B^2$ is presented by Fig. 1.
\begin{figure}[h]
\includegraphics[height=4.0in,width=4.0in]{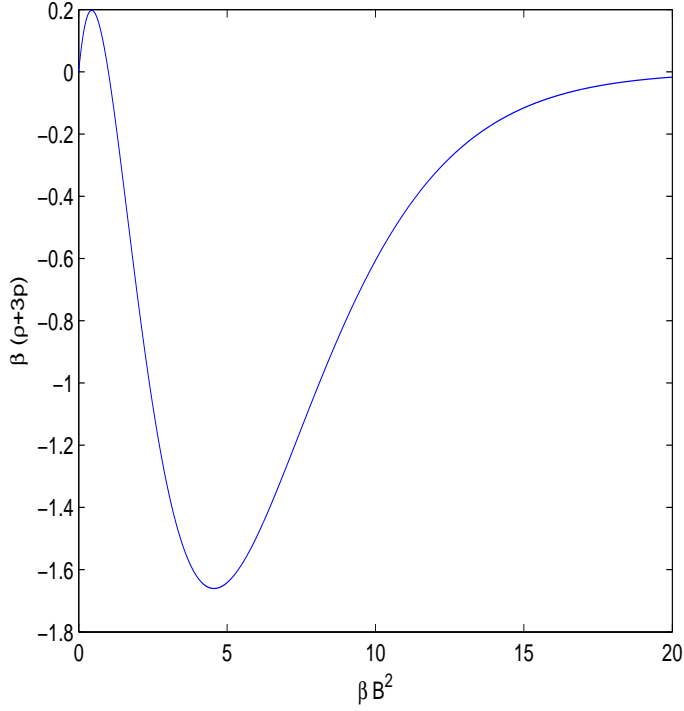}
\caption{\label{fig.1}The function  $\beta(\rho+3p)$ vs. $\beta B^2$. }
\end{figure}
The acceleration of the universe, $\rho + 3p < 0$, takes place at $\beta B^2>1$ (see Fig. 1). In the early universe the magnetic field is strong and is the source of inflation.
The energy-momentum tensor conservation, $\nabla^\mu T_{\mu\nu}=0$, (for FRW metric (19)) leads to the relation
\begin{equation}
\dot{\rho}+3\frac{\dot{a}}{a}\left(\rho+p\right)=0.
\label{25}
\end{equation}
From Eqs. (21),(22), for the case $\textbf{E} = 0$, one finds
\begin{equation}
\rho=\frac{B^2}{2}\exp(-\beta B^2/2),~~~~\rho+p=\frac{2}{3}B^2\left(1-\frac{\beta B^2}{2}\right)\exp(-\beta B^2/2).
\label{26}
\end{equation}
After integrating Eq. (25), taking into consideration Eqs. (26), we obtain the solution
\begin{equation}
B(t)=\frac{B_0}{a^2(t)},
\label{27}
\end{equation}
where $B_0$ is the present value of the magnetic field $B$ and $a(t)=1$ today. In the early universe $a\rightarrow 0$. Due to inflation the scale factor $a(t)$ increases and the magnetic field decreases. At the weak magnetic field NLED becomes Maxwell's electrodynamics at the present time. The energy density and pressure (at $\textbf{E}=0$) as the function of the scale factor are given by
\[
\rho(t)=\frac{B_0^2}{2a^4(t)}\exp\left(-\beta B_0^2/2a^4(t)\right),
\]
\begin{equation}
p(t)=\frac{B_0^2}{6a^4(t)}\left(1-\frac{2\beta B_0^2}{a^4(t)}\right)\exp\left(-\beta B_0^2/2a^4(t)\right).
\label{28}
\end{equation}
From Eq. (28) one finds
\begin{equation}
\lim_{a(t)\rightarrow 0}\rho(t)=\lim_{a(t)\rightarrow 0}p(t)=0,~~\lim_{a(t)\rightarrow \infty}\rho(t)=\lim_{a(t)\rightarrow \infty}p(t)=0.
\label{29}
\end{equation}
Thus, the singularities of the energy density and pressure at $a(t)\rightarrow 0$ and $a(t)\rightarrow \infty$ are absent.
From Eqs. (28) we obtain the equation of state (EoS), $w(t)=p(t)/\rho(t)$,
\begin{equation}
w(t)=\frac{1}{3}\left(1-\frac{2\beta B_0^2}{a^4(t)}\right).
\label{29}
\end{equation}
At $a(t)\rightarrow \infty$, we have $w(t)\rightarrow 1/3$, i.e. EoS for ultra-relativistic case \cite{Landau}.
From the Einstein equation (17) we obtain the Ricci scalar expressed through the energy-momentum tensor trace (5),
\begin{equation}
R=\kappa^2T_{\mu}^{~\mu}=\kappa^2(\rho-3p)=\frac{\kappa^2\beta B_0^4}{a^8(t)}\exp(-\beta B_0^2/a^4(t)).
\label{31}
\end{equation}
From Eq. (31) one finds
\begin{equation}
\lim_{a(t)\rightarrow 0}R(t)=\lim_{a(t)\rightarrow \infty}R(t)=0.
\label{32}
\end{equation}
As a result, there are not singularities of the curvature at early and late times of the universe evolution. From FRW metric one can obtain the Ricci tensor squared, $R_{\mu\nu}R^{\mu\nu}=\kappa^4(\rho^2+3p^2)$, which has no singularities at $a(t)\rightarrow 0$ and $a(t)\rightarrow \infty$. The Kretschmann scalar $R_{\mu\nu\alpha\beta}R^{\mu\nu\alpha\beta}$ can be expressed via linear combinations of $\kappa^4\rho^2$, $\kappa^4\rho p$, and $\kappa^4p^2$ \cite{Kruglov3}. Therefore, it also does not possess singularities at $a(t)\rightarrow 0$ and $a(t)\rightarrow \infty$.

\section{The evolution of the universe}

To find the dependence of the scale factor on the time we explore the second Friedmann equation for three dimensional flat universe
\begin{equation}
\left(\frac{\dot{a}}{a}\right)^2=\frac{\kappa^2\rho}{3}.
\label{33}
\end{equation}
Replacing the energy density from Eq. (28) into Eq. (33) we obtain the equation as follows:
\begin{equation}
\dot{a}^2 =\frac{\kappa^2B_0^2}{6a^2(t)}\exp\left(-\beta B_0^2/2a^4(t)\right).
\label{34}
\end{equation}
Integrating Eq. (34) one finds
\begin{equation}
t=\frac{\sqrt{6}}{\kappa B_0}\int a\exp\left(\beta B_0^2/(4a^4)\right)da.
\label{35}
\end{equation}
Eq. (35) can be represented as
\begin{equation}
y=\int xdx\exp\left(1/x^4\right)=\frac{1}{2}\left(x^2\exp\left(\frac{1}{x^4}\right)-\sqrt{\pi}\mbox{erfi}
\left(\frac{1}{x^2}\right)\right)+C,
\label{36}
\end{equation}
where
\begin{equation}
y=\kappa t\sqrt{\frac{2}{3\beta}},~~x=a(t)\left(\frac{4}{\beta B_0^2}\right)^{1/4},~~\mbox{erfi}(x)=\frac{2}{\sqrt{\pi}}\int_0^x \exp(t^2)dt,
\label{37}
\end{equation}
$\mbox{erfi}(x)$ is the imaginary error function.
Without loss of generality the constant of integration $C$ can be omitted, $C=0$, because it gives only the shift of the cosmic time. The plot of the function $y(x)$ is given in Fig. 2.
\begin{figure}[h]
\includegraphics[height=4.0in,width=4.0in]{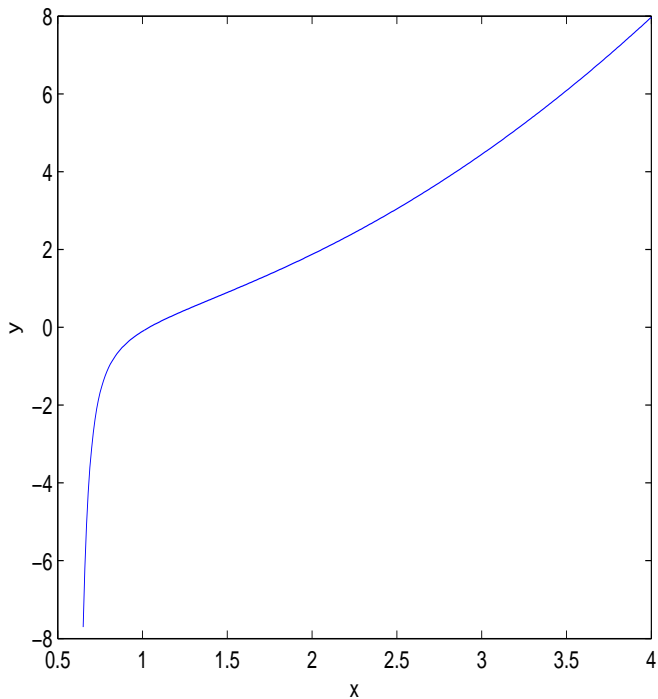}
\caption{\label{fig.2}The function  $y=\kappa t\sqrt{\frac{2}{3\beta}}$ vs. $x=a(t)\left(\frac{4}{\beta B_0^2}\right)^{1/4}$.}
\end{figure}
From Eq. (36) we find the approximate solution at $x\rightarrow\infty$ ($a(t)\rightarrow\infty$)
\begin{equation}
y=\frac{x^2}{2}-\frac{1}{2x^2}+{\cal O}\left(\left(\frac{1}{x^6}\right)\right).
\label{38}
\end{equation}
Eq. (38) at $t\rightarrow\infty$ leads to
\begin{equation}
a\rightarrow \sqrt{\kappa B_0 t}\left(\frac{2}{3}\right)^{1/4}.
\label{39}
\end{equation}
Thus, the solution (39) for large time $t$ is $a\propto \sqrt{t}$ and it corresponds to the radiation era.
From Eq. (36) we obtain another limiting case at $x\rightarrow 0$,
\begin{equation}
y=-e^{1/x^4}\left(\frac{1}{4}x^6+\frac{3}{8}x^{10}+{\cal O}\left(x^{12}\right)\right).
\label{40}
\end{equation}
Eq. (40) shows that at $t\rightarrow -\infty$ the scale factor $a(t)\rightarrow 0$. It means that the size of the universe, $a(t)$,  never was zero. Numerical calculations give $x=1.04023$ at $y=0$ (or $t=0$). Thus, at $t=0$, $a(0)(4/\beta B_0^2)^{1/4}=1.04023$ or $a(0)\approx 0.7356\beta^{1/4}\sqrt{B_0}$.
As a result, there are no singularity for the scale factor and the universe accelerates at $a(t)<\beta^{1/4}\sqrt{B_0}$. Then, at $a(t)>\beta^{1/4}\sqrt{B_0}$ the universe undergoes the deceleration. The model describes inflation without singularities and there are no problems with initial conditions.

\subsection{Causality, classical stability and speed of sound}

The causality occurs if the speed of the sound is less that the light speed, $c_s\leq 1$ \cite{Quiros}. A classical stability requires that the square sound speed is positive, i.e. $c^2_s\geq 0$.
The sound speed is defined from the equation describing the evolution of linear adiabatic perturbations of the background energy density, $\rho_B(t,\textbf{r})=\rho_B(t)+\delta\rho_B(t,\textbf{r})$. The wave equation for small perturbations follows from the conservation of the energy-momentum tensor, $\delta\ddot{\rho}_B=c_s^2\nabla^2\delta\rho_B$ \cite{Peebles}.
Thus, the bound $c_s^2\geq 0$ is required on the speed of sound to have small perturbations of the background energy density $\delta\rho_B(t,\textbf{r})$. Otherwise, the energy density perturbations will grow and there will be a classical instability. From Eqs. (21), (22) we obtain at $E=0$  the sound speed squared
\begin{equation}
c^2_s=\frac{dp}{d\rho}=\frac{dp/d{\cal F}}{d\rho/d{\cal F}}=\frac{1-9\beta B^2/2+\beta^2B^4}{3\left(1-\beta B^2/2\right)}.
\label{41}
\end{equation}
The classical stability, $c^2_s\geq 0$, takes place at $(9+\sqrt{65})/4\geq\beta B(t)\geq2$ or at $(9-\sqrt{65})/4\geq\beta B^2(t)>0$. The bound $c_s\leq 1$, required for the causality, occurs at $(3-\sqrt{17})/2\leq\beta B^2(t)\leq 2$ or at $\beta B^2(t)\geq(3+\sqrt{17})/2$.
Both requirements $c^2_s\geq 0$, $c_s\leq 1$ lead to $(9+\sqrt{65})/4\geq\beta B^2(t)\geq (3+\sqrt{17})/2$ or $\beta B^2(t)\leq (9-\sqrt{65)}/4$. From Eq. (27) we obtain the range where $1\geq c^2_s\geq 0$
\begin{equation}
\sqrt[4]{\frac{2\beta B_0^2}{3+\sqrt{17}}}\geq a(t)\geq\sqrt[4]{\frac{4\beta B_0^2}{9+\sqrt{65}}} ~~~\mbox{or}~~~a(t)\geq\sqrt[4]{\frac{4\beta B^2_0}{9-\sqrt{65}}}.
\label{42}
\end{equation}
We have the acceleration of the universe at $a(t)<\sqrt[4]{\beta B_0^2}$. Thus, the causality and a classical stability take place if $0.728\sqrt[4]{\beta B_0^2}\geq a(t)\geq 0.696\sqrt[4]{\beta B_0^2}$ corresponding to inflation phase of the universe, or at $a(t)\geq 1.437\sqrt[4]{\beta B_0^2}$ corresponding to the deceleration of the universe. At $a\rightarrow\infty$ the sound speed squared is $c^2_s\rightarrow 1/3$. The plot of $c^2_s$ vs. $a/(\beta B_0^2)^{1/4}$ is given in Fig. 3 for the inflation phase.
\begin{figure}[h]
\includegraphics[height=4.0in,width=4.0in]{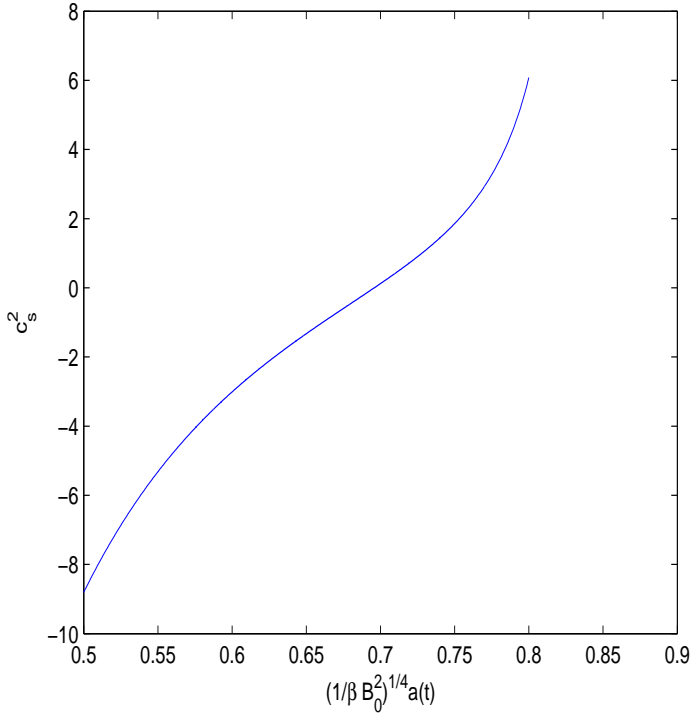}
\caption{\label{fig.3} The function $c^2_s$ vs. $a/(\beta B_0^2)^{1/4}$.}
\end{figure}
It should be stressed that the condition $c^2_s > 1$ does not necessarily lead to a violation of
causality. One way of proving causality is to verify that the field equations for
the electromagnetic field remain hyperbolic \cite{Uzan}, \cite{Golovnev}.

\subsection{Cosmological parameters}

One finds from Eq. (21) (at $E=0$) the relation $\beta {\cal F}=-W(-\beta\rho)$, where $W(x)$ is the Lambert function. Then from Eq. (22) we obtain  the equation
\begin{equation}
p=-\rho+f(\rho),~~~~f(\rho)=\frac{4}{3}\rho\left[1+W(-\beta\rho\right].
\label{43}
\end{equation}
representing EoS for the perfect fluid.
When the condition $|f(\rho)/\rho|\ll 1$ holds during inflation, one can obtain the expressions for the spectral index $n_s$, the tensor-to-scalar ratio $r$, and the running of the spectral index $\alpha_s=dn_s/d\ln k$ \cite{Odintsov}
\begin{equation}
n_s\approx 1-6\frac{f(\rho)}{\rho},~~~r\approx 24\frac{f(\rho)}{\rho},~~~\alpha_s\approx -9\left(\frac{f(\rho)}{\rho}\right)^2.
\label{44}
\end{equation}
The inequality $|f(\rho)/\rho|\ll 1$ leads to $W_0(-\beta\rho)\ll -1/4$ ($W_0(x)$ corresponds to the upper branch of the Lambert function) or $\beta\rho\gg 1/(4e^{1/4})\approx0.1947$. In terms of the scale factor this condition reads $a\ll(2\beta B_0^2)^{1/4}$. For the acceleration phase of the universe ($a<(\beta B_0^2)^{1/4}$) this requirement can be satisfied. Therefore, the parameters (44) may be fulfilled in the inflation phase. From Eqs. (43),(44) we find the relations
\begin{equation}
r=4(1-n_s)=8\sqrt{-\alpha_s}=32\left[1+W_0(-\rho\beta)\right].
\label{45}
\end{equation}
The PLANCK \cite{Ade} and WMAP \cite{Komatsu}, \cite{Hinshaw} data are
\[
n_s=0.9603\pm 0.0073 ~(68\% CL),~~~r<0.11 ~(95\%CL),
\]
\begin{equation}
\alpha_s=-0.0134\pm0.0090 ~(68\% CL).
\label{46}
\end{equation}
The BICEP2 experiment \cite{Ade1} for the tensor-to-scalar ratio has the value $r=0.20^{+0.07}_{-0.05}~(68\% CL)$, and the validity of this value was challenged. One can take the value $r=0.13$ and from Eqs. (45) we obtain the cosmological parameters: $n_s=0.9675$,  $\alpha_s=-2.64\times 10^{-4}$. From Eq. (45) one can find $\beta\rho=0.367876$ ($\beta{\cal F}=0.9959$) or $a\approx 0.8418(\beta B_0^2)^{1/4}$ corresponding to the inflation phase. The plots of the functions $n_s$, $r$ and $\alpha_s$ vs. $a/(\beta B_0^2)^{1/4}$ are represented in Fig. 4.
\begin{figure}[h]
\includegraphics[height=4.0in,width=4.0in]{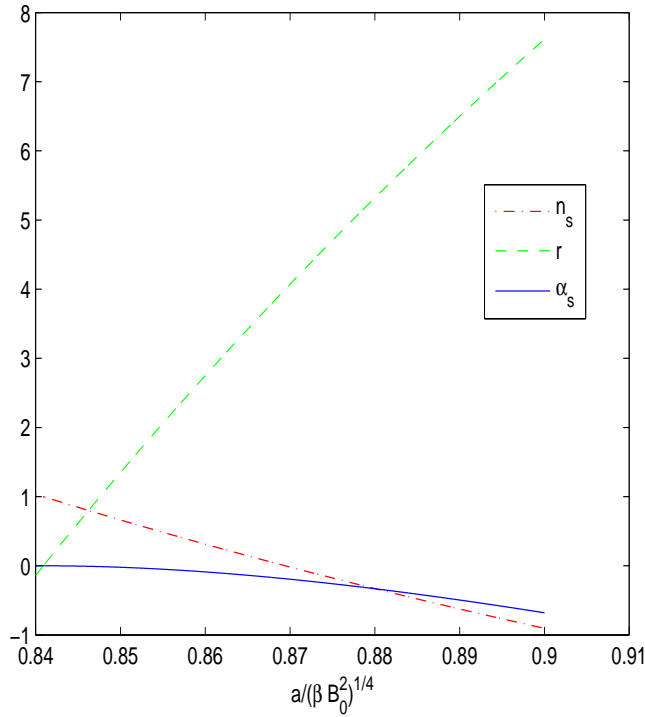}
\caption{\label{fig.4} The functions $n_s$, $r$ and $\alpha_s$ vs. $a/(\beta B_0^2)^{1/4}$.}
\end{figure}
Thus, we have obtained the reasonable cosmological parameters.

\section{Conclusion}

A new model of nonlinear electromagnetic fields with the dimensional parameter $\beta$ is proposed. The scale invariance and dual invariance are broken because the trace of the energy-momentum tensor is not zero. From PVLAS experiment the bound on the parameter $\beta$ was estimated. We consider the nonlinear electromagnetic fields coupled to the gravitation field. The magnetic stochastic background in FRW spacetime is the source of the universe inflation. The origin of large scale magnetic fields is one of the problems in modern cosmology. It follows from Eq. (27) that supporting magnetic fields in cosmology on large scales are weak. The mechanism proposed in [13] can explain the galactic and intergalactic magnetic fields by a hypothesis of the existence of millicharged dark matter particles. Probably there are other mechanisms that make clear the long standing problem of the galactic and intergalactic magnetic fields.

The universe accelerates at $a(t)<\beta^{1/4} \sqrt{B_0}$, where $B_0$ is the magnetic induction field at the present time and $\beta$ is a free parameter in the model which can be found from experimental data. We show that there are no singularities in the energy density, pressure, the Ricci scalar and Ricci tensor squired. At $a(t)>\beta^{1/4} \sqrt{B_0}$ the universe undergoes the deceleration and the scale factor approaches $a(t)\propto \sqrt{t}$ at $t\rightarrow \infty$ that corresponds to the radiation era. We have obtained the range of the scale factor when the causality of the model and a classical stability take place. The spectral index, the tensor-to-scalar ratio, and the running of the spectral index estimated are in the approximate agreement with the PLANCK, WMAP, and BICEP2 data.
It was demonstrated, in the framework of inflationary cosmology, that it is possible to describe the universe inflation without introduction dark energy, the cosmological constant and modification of GR.
Thus, the model of NLED proposed can describe inflation of the universe.

There are some similarities and differences between BI model and NLED proposed. The duality symmetry holds in BI model but in our model the duality symmetry is broken. The duality symmetry is also violated in QED where the effective Heisenberg-Euler Lagrangian is induced by quantum corrections.
The birefringence phenomenon takes place in NLED proposed and in QED with quantum corrections \cite{Kruglov2} but in BI model there is no the birefringence effect. In our model nonlinear electromagnetic fields drives the universe to accelerate and no such important effect in BI electrodynamics coupled to GR \cite{Novello1}. Therefore, the model proposed is of definite theoretical interest. In addition, BI model suffers a serious causality problems \cite{Quiros}.

The coupling of electromagnetic Lagrangian with the $F(R)$ term in modified gravity was investigated in
\cite{Nojiri}, \cite{Nojiri1} (see also\cite{Cruz}). It was demonstrated the generation of large-scale magnetic fields due to the breaking of the conformal invariance of the electromagnetic field through its non-minimal gravitational coupling. It is interesting to study the current model with the additional non-minimal term $F(R)F_{\mu\nu}F^{\mu\nu}$. One can expect the similar effects of the generation of large-scale magnetic fields in the NLED proposed coupling with gravity. Probably the
late-time acceleration of the universe also can be realized in such extension of the model.
We leave such study for further investigation.

\end{document}